%% file: resonancePRL.tex
\documentclass[aps,prl,reprint,superscriptaddress,showpacs,preprintnumbers]{revtex4-1}

\usepackage{graphicx}

\usepackage{amssymb,amssymb}

\usepackage{lineno}

\usepackage{hyperref}
\usepackage{subfigure}

\newlength{\figwidth}
\def\integLumi{315~nb$^{-1}$}
\def\integLumiUncertainty{$\pm 11$\%}
\def\JERuncertainty{14} 
\def\MjjCut{200} 
\def\MjjLowerExclusion{300} 
\def\MjjLowerExclusionTeV{0.30} 
\def\MjjUpperFitRange{1900} 
\def\MjjLowerLimitMRSTTeV{1.26} 
\def\MjjLowerLimitMRSThalfmuRmuFTeV{1.32} 
\def\MjjLowerLimitMRSTtwicemuRmuFTeV{1.22} 
\def\MjjLowerLimitMRSTEXPECTEDTeV{1.06} 
\def\MjjLowerLimitStatOnlyMRSTTeV{1.28} 
\def\MjjLowerLimitCTEQsixLoneTeV{1.20} 
\def\MjjLowerLimitCTEQsixLoneEXPECTEDTeV{0.99} 
\def\MjjLowerLimitStatOnlyCTEQsixLoneTeV{1.23} 
\def\MjjLowerLimitCTEQfiveLTeV{1.22} 
\def\MjjLowerLimitCTEQfiveLEXPECTEDTeV{1.00} 
\def\MjjLowerLimitStatOnlyCTEQfiveLTeV{1.25} 
\def\NpseudoExperiments{$10^3$}

\def\triggerThreshold{15} 
\def\pTcut{80} 
\def\topbottomLabel{~}
\def\GeVcc{${\rm GeV}$}
\def\TeVcc{${\rm TeV}$}
\def\GeVc{${\rm GeV}$}

\begin{document}


\preprint{CERN-PH-EP-2010-028}

\title{Search for New Particles in Two-Jet Final States
in 7~TeV Proton-Proton Collisions with the ATLAS Detector at the LHC}


\author{The ATLAS Collaboration}


\date{September 6, 2010}

\begin{abstract}
  A search for new heavy particles manifested as resonances in
two-jet final states is presented.  The data were produced in 7~TeV
proton-proton collisions by the Large Hadron Collider (LHC) and
correspond to an integrated luminosity of \integLumi~collected by the
ATLAS detector.  No resonances were observed.  Upper limits were
set on the product of cross section and signal acceptance for
excited-quark ($q^*$) production as a function of $q^*$ mass.  These
exclude at the 95\%~CL the $q^*$ mass interval $\MjjLowerExclusionTeV
< m_{q^*} < \MjjLowerLimitMRSTTeV$~\TeVcc, extending the reach of
previous experiments.
  \end{abstract}


\maketitle

Two-jet (dijet) events in high-energy proton-proton ($pp$) collisions
are usually described in the Standard Model (SM) by applying quantum
chromodynamics (QCD) to the scattering of beam-constituent quarks and
gluons.  Several extensions beyond the SM predict new heavy particles,
accessible at LHC energies, that decay into two energetic partons.
Such new states may include an excited composite quark $q^*$,
exemplifying quark
substructure~\cite{Eichten:1983hw,Baur:1987ga,Baur:1989kv}; an
axigluon predicted by chiral color
models~\cite{Frampton:1987dn,Bagger:1987fz}; a flavour-universal
color-octet coloron~\cite{Chivukula:1996yr,Simmons:1996fz}; or a
color-octet techni-$\rho$ meson
predicted by models of extended technicolor and topcolor-assisted
technicolor~\cite{Lane:1991qh,Lane:2002sm,Foadi:2007ue,Belyaev:2008yj}.

Particularly sensitive to such new objects is the dijet invariant mass
observable, defined as $m^{jj} \equiv \sqrt{(E^{ j_1} + E^{j_2})^2 -
(\vec{p}^{j_1} + \vec{p}^{j_2})^2}$, where $E$ and $\vec{p}$ are the
jet energy and momentum, respectively.  Several experiments have
examined $m^{jj}$ distributions in search of new
resonances~\cite{Alitti19933,Abe:1995jz,PhysRevD.55.R5263,PhysRevD.69.111101,CDF_Mjj,Aaron2009335};
recently, 1.13~fb$^{-1}$ of $p\bar{p}$ collision data at the Fermilab
Tevatron collider have excluded the existence of excited quarks $q^*$
with mass $260 < m_{q^*} < 870$~\GeVcc~\cite{CDF_Mjj}.  This Letter
reports the first search by the ATLAS experiment~\cite{DetectorPaper}
at the LHC for such massive particles in $pp$ collisions at a
centre-of-mass energy of $\sqrt{s} = 7$~TeV, based on a data sample
corresponding to an integrated luminosity of \integLumi.  The analysis
presented here focused on a search for excited quarks because of the
accessible predicted cross section~\cite{Baur:1987ga,Baur:1989kv}\ for
such particles and the benchmark nature of the model that allows
limits on acceptance times cross section to be set for resonant states
with intrinsic widths narrower than the experimental resolution.

The analysis technique consisted of a model-independent search for a
dijet mass resonance on top of a smooth and rapidly falling spectrum
and relied on the measured $m^{jj}$ distribution to estimate the
background level to this new possible signal.  In the absence of an
observed new physics signal, upper limits were determined on products
of cross section~($\sigma$) and signal acceptance~(${\cal A}$) for
several $q^*$ test masses for a standard set of model parameters.

The ATLAS detector~\cite{DetectorPaper} is a multipurpose particle
physics apparatus with a forward-backward symmetric cylindrical
geometry and near 4$\pi$ coverage in solid angle~\footnote{The ATLAS
reference system is a Cartesian right-handed coordinate system, with
the nominal collision point at the origin.  The anticlockwise beam
direction defines the positive $z$-axis, while the positive $x$-axis
points from the collision point to the centre of the LHC ring and the
positive $y$-axis points upward.  The angles $\phi$ and $\theta$\ are
the azimuthal and polar angles.  The pseudorapidity is defined as
$\eta = -\ln\left[\tan\left(\theta / 2\right)\right]$\ and rapidity is
defined as $y = \frac{1}{2}\ln\left[(E+p_z)/(E-p_z)\right]$, where $E$
is the energy and $p_z$ is the longitudinal component of the momentum
along the beam direction.}.  The overall layout of the detector is
dominated by its four superconducting magnet systems, which comprise a
thin solenoid surrounding inner tracking detectors and three large
toroids with an eightfold azimuthal symmetry.

The calorimeters, which are surrounded by an extensive muon system,
are of particular importance to this analysis.  In the pseudorapidity
region $\left|\eta\right| < 3.2$, high-granularity liquid-argon (LAr)
electromagnetic sampling calorimeters are used.  An iron-scintillator
tile calorimeter provides hadronic coverage in the range
$\left|\eta\right| < 1.7$.  The end-cap and forward regions, spanning
$1.5 < \left|\eta\right| < 4.9$, are instrumented with LAr calorimetry
for both electromagnetic and hadronic measurements.



The data sample was collected during
stable periods of 7~TeV $pp$ collisions using a
trigger configuration requiring the lowest-level
hardware-based calorimeter jet trigger to satisfy a nominal
transverse energy threshold
of \triggerThreshold~GeV~\cite{jetProdICHEP}.  This trigger had an
efficiency greater than 99\% for events with at least one jet with transverse
energy higher than 80~GeV.

Jets were reconstructed using the anti-$k_T$ jet clustering
algorithm~\cite{antikT} with a radius parameter $R = 0.6$.  The inputs
to this algorithm were clusters of calorimeter cells
seeded by cells with energy significantly above the measured noise.
Jet four-vectors were constructed by performing a four-vector sum over
these cell clusters, treating each as an ($E,\, \vec{p}$) four-vector
with zero mass.  These were corrected for the effects of calorimeter
non-compensation and inhomogeneities by using transverse-momentum
($p_T$) and $\eta$-dependent calibration factors based on Monte Carlo
(MC) corrections and validated with extensive test-beam and
collision-data studies~\cite{CombinedTB,jetProdICHEP}.  The $m^{jj}$
observable was computed without unfolding jets to hadrons or partons.

In order to suppress cosmic-ray and beam-related backgrounds, events
were required to contain at least one primary collision vertex,
defined by at least five reconstructed charged-particle tracks, each
with a position, when extrapolated to the beamline, of $\left|z\right|
< 10$~cm.  Events with at least two jets were retained if the highest
$p_T$\ jet (the ``leading'' jet) satisfied $p_T^{j_1} > \pTcut$~\GeVc\
and the next-to-leading jet satisfied $p_T^{j_2} > 30$~\GeVc; this
ensured that the data sample had high and unbiased trigger and jet
reconstruction efficiencies.  Those events containing a poorly
measured jet with $p_T > 15$~\GeVc\ were vetoed to prevent cases where
a jet was incorrectly identified as one of the two leading
jets~\cite{JetEtMiss_Cleaning_Note}; this affected the event selection
by less than 0.5\%.  The two leading jets were required to satisfy
several quality criteria~\cite{JetEtMiss_Cleaning_Note} and to lie
outside detector regions where the jet energy was not yet measured in
an optimal way, such as the interval $1.3 < |\eta^{\rm jet} | < 1.8$.
Finally, both jets were required to be in the pseudorapidity region
$\left|\eta^{\rm jet}\right| < 2.5$, and their pseudorapidity difference was
required to satisfy $\left|\eta^{j_1}-\eta^{j_2}\right| < 1.3$.  These
cuts, which suppress high-mass SM multijet background, were
determined by performing an optimization of the potential signal from
$q^*$\ decays (using a $q^*$~mass of 1~\TeVcc) compared with the SM
background.  There were 132,433 candidates that satisfied these
requirements.

\begin{figure}
\begin{center}
\includegraphics[width=0.5\figwidth]{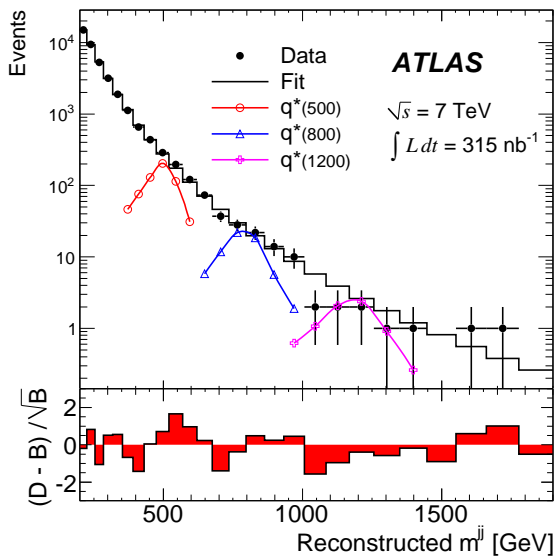}
\end{center}
\caption{ \topbottomLabel The data (D) dijet mass distribution
(filled points) fitted using a binned background (B) distribution described
by Eqn.~\ref{eq:f} (histogram).  The predicted $q^*$
signals~\cite{Baur:1987ga,Baur:1989kv} for excited-quark masses of
500, 800, and 1200~\GeVcc\ are overlaid, and the bin-by-bin
significance of the data-background difference is shown.}
\label{fig:spectrumAndFit}
\end{figure}

The final event sample was selected by requiring the dijet invariant
mass to satisfy $m^{jj} > \MjjCut$~\GeVcc\ in order to eliminate any
potential kinematic bias in the $m^{jj}$\ distributions from the
selection requirements on the jet candidates.  There were 37,805
events in this sample, which formed the $m^{jj}$ distribution shown in
Fig.~\ref{fig:spectrumAndFit}.

MC signal events were generated using the excited-quark
($qg\to q^*$) production model~\cite{Baur:1987ga,Baur:1989kv}.
The excited quark $q^*$ was assumed to have spin 1/2 and quark-like
couplings, relative to those of the SM $SU(2)$, $U(1)$, and $SU(3)$
gauge groups, of $f=f^\prime=f_s=1$, respectively.  The compositeness
scale ($\Lambda$) was set to the $q^*$ mass.  Signal events were
generated using {\sc pythia}~\cite{pythia} 6.4.21, a leading-order
parton-shower MC generator, with the modified leading-order
MRST2007~\cite{Sherstnev:2007nd} parton distribution functions (PDFs)
and with the renormalization and factorization scales set to the mean
$p_T$ of the two leading jets.  {\sc pythia} was also used to decay
the excited quarks to all possible SM final states, which were
dominantly $qg$ but also $qW$, $qZ$, and $q\gamma$.  The MC samples
were produced~\cite{ATLSIM} using the ATLAS MC09 parameter
tune~\cite{ATLAS_MC09} and a {\sc geant4}-based detector
simulation~\cite{Agostinelli:2002hh}.

Figure~\ref{fig:spectrumAndFit} shows the predicted signals
for $q^*$ masses of 500, 800, and 1200~\GeVcc, after all selection
cuts.  The signal acceptance (${\cal A}$), which included
reconstruction and trigger efficiencies near 100\%, was found to range
from $\sim\!31$\% for $m_{q^*} = 300$~\GeVcc\ to $\sim\!48$\% for
$m_{q^*} = 1.7$~\TeVcc~\footnote{For the specific $qg$ final state,
the product of branching fraction and acceptance ranged from
$\sim\!28$\% to $\sim\!40$\% for $m_{q^*} = 300$~\GeVcc\ and $m_{q^*}
= 1.7$~\TeVcc, respectively.}.  The choice of dijet mass binning was
motivated by the dijet mass resolution of the signal.  The predicted
experimental width ranged from $\sigma_{m^{jj}}/m^{jj} \sim 11$\% at
$m_{q^*} = 300$~\GeVcc\ to $\sigma_{m^{jj}}/m^{jj} \sim 7$\% at
$m_{q^*} = 1.7$~\TeVcc\ and was dominated by the
detector energy resolution.

The background shape was determined by fitting the observed spectrum with
the function~\cite{CDF_Mjj}
\begin{equation}
f(x) = p_1 (1 - x)^{p_2} x^{p_3 + p_4\ln x},
\label{eq:f}
\end{equation}
where $x\equiv m^{jj}/\sqrt{s}$ such that $f(1)=0$ and $f(0)\to
+\infty$, and $p_{\{1,2,3,4\}}$ are free parameters.  The $x^{p_4\ln
x}$ factor was included to describe the high-$m^{jj}$ part of the
spectrum.  The function in Eqn.~\ref{eq:f} has been shown to fit the
$m^{jj}$ observable well in {\sc pythia}, {\sc herwig}, and
next-to-leading-order (NLO) perturbative QCD predictions for
$p\bar{p}$ collisions at $\sqrt{s} = 1.96$~TeV~\cite{CDF_Mjj}.
Studies using {\sc pythia} and the ATLAS {\sc
geant4}-based detector simulation
were performed to
demonstrate that the smooth and monotonic form of Eqn.~\ref{eq:f}
describes QCD-predicted dijet mass distributions in $pp$\ collisions
at $\sqrt{s} = 7$~TeV.
There is good agreement between the MC prediction and the fitted parametrization
in Eqn.~\ref{eq:f},
as evidenced by a
$\chi^2$ per degree of freedom of 27/22 over the dijet mass range $200 < m^{jj} <
\MjjUpperFitRange$~\GeVcc.

The results of fitting the data with Eqn.~\ref{eq:f}\ are shown in
Fig.~\ref{fig:spectrumAndFit}.
The presence or absence of detectable $m^{jj}$ resonances in
this distribution was determined by performing several statistical
tests of the background-only hypothesis.
A suite of six tests was employed: 
the BumpHunter~\cite{Aaltonen:2008vt}, 
the Jeffreys divergence~\cite{Jeffreys}, 
the Kolmogorov-Smirnov test, 
the likelihood,
the Pearson $\chi^2$, 
and the TailHunter statistic~\cite{Aaltonen:2007dg}. 
The agreement of the data with the background-only hypothesis of 
a smoothly varying and monotonic distribution
was determined for each statistic by calculating the $p$-value for the
data using \NpseudoExperiments~pseudo-spectra drawn from Poisson
variations seeded by the results of the fit of Eqn.~\ref{eq:f} to the
data.
The $p$-value of the background-only hypothesis is defined as the
fraction of pseudo-experiments that result in a value of the given
statistic greater than the value of the same statistic found by the
fit to the data.  The results of all six tests were consistent with
the conclusion that the fitted parametrization described the observed
data distribution well, with $p$-values
in excess of 51\%.
These observations supported the background-only hypothesis.  

In the absence of any observed discrepancy with the zero-signal
hypothesis, a Bayesian approach was used to set 95\% credibility-level
(CL) upper limits on $\sigma\cdot {\cal A}$ for hypothetical new
particles decaying into dijets with $\left|\eta^{\rm jet}\right| <
2.5$.  For each of the test masses (indexed by $\nu$) corresponding to
the excited-quark $q^*$ predictions, a likelihood function $L_\nu$ was
defined as a product of Poisson factors computed for each bin ($i$) of
the $m^{jj}$ distribution:
\begin{equation}
\label{eq:Likelihood}
L_\nu(d | b_\nu,s) \equiv
\prod_i \frac{\left[b_{\nu i}+s_{i}(\nu)\right]^{d_i}}{d_i!}
          e^{-\left[b_{\nu i}+s_{i}(\nu)\right]},
\end{equation}
where $d_i$ is the observed number of data events in bin $i$, $b_{\nu i}$ is the
background in bin $i$ obtained as described below,
and $s_i(\nu)$ is the predicted signal added in bin $i$ by the signal
template; the latter was normalized to the total number of predicted signal
events $s = \sum_i{s_i(\nu)}$.  For each $\nu$, the backgrounds in the
bins $b_{\nu i}$ were evaluated from a simultaneous five-parameter fit
of the signal and background distributions to ensure that the
background determination would not be biased by the presence of any
signal.  The four background parameters were those in Eqn.~\ref{eq:f};
the fifth parameter consisted of the normalization of the predicted
$\nu^{\rm th}$ $q^*$ signal template.  
To avoid acceptance bias, the lowest $q^*$ test
mass used was \MjjLowerExclusion~GeV.
For every $q^*$\ mass, Eqn.~\ref{eq:Likelihood} was computed for a
range of possible signal yields, $s$, and the resulting likelihood
function was multiplied by a flat prior in $s$\ to give a posterior
probability density in $s$.  The 95\% probability region was then
determined by integration of the posterior probability distribution.
This Bayesian technique was found to yield credibility intervals that
corresponded well with frequentist confidence intervals.  This was
verified by performing a series of pseudo-experiments to determine, by
way of a standard frequentist calculation, the coverage, or the
fraction of times that the 95\% Bayesian credibility interval
contained the true number of signal events.

The dominant sources of systematic uncertainty, in decreasing order of
importance, were the absolute jet energy scale (JES), the background
fit parameters, the integrated luminosity, and the jet energy
resolution (JER).  The JES uncertainty was quantified as a function of
$p_T$ and $\eta^{\rm jet}$, with values in the range
$6\sim9\%$~\cite{jetProdICHEP,JES,JES2}.  The jet calibration relied
on the MC simulation of the response of the ATLAS detector; its
uncertainty was constrained by varying the ATLAS simulation and from
{\it in situ} information.
The systematic uncertainty on the determination of the background was
taken from the uncertainty on the parameters resulting from the fit of
Eqn.~\ref{eq:f} to the data sample.  The uncertainty on $\sigma\cdot
{\cal A}$ due to integrated luminosity was estimated to be
\integLumiUncertainty~\cite{ATLAS_lumi}.  The JER uncertainty was
treated as uniform in $p_T$ and $\eta^{\rm jet}$ 
with a value of $\pm
\JERuncertainty$\% on the fractional $p_T$ resolution of each
jet~\cite{JER}.
The effects of JES, background fit, integrated luminosity, and JER
were incorporated as nuisance parameters into the likelihood function
in Eqn.~\ref{eq:Likelihood} and then marginalized by numerically
integrating the product of this modified likelihood, the prior in $s$,
and the priors corresponding to the nuisance parameters to arrive at a
modified posterior probability distribution.  In the course of
applying this convolution technique, the JER was found to make a
negligible contribution to the overall systematic uncertainty.

\begin{figure}
\begin{center}
\includegraphics[width=0.5\figwidth]{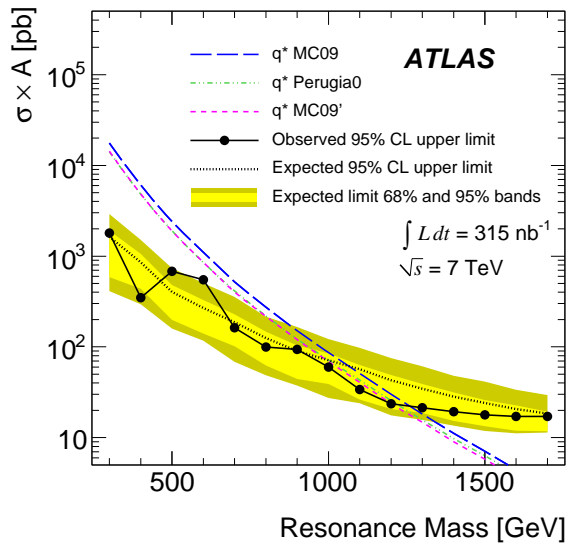}
\end{center}
\caption{The 95\% CL upper limit on $\sigma\cdot {\cal A}$ as a
function of dijet resonance mass (black filled circles).  The black
dotted curve shows the expected 95\% CL upper limit and the light and
dark yellow shaded bands represent the 68\% and 95\% credibility
intervals of the expected limit, respectively.  The dashed curves
represent excited-quark $\sigma\cdot {\cal A}$ predictions for
different MC tunes, each using a different PDF set.}
\label{fig:ExclusionWithStatSystematics}
\end{figure}

Figure~\ref{fig:ExclusionWithStatSystematics} depicts the resulting
95\% CL upper limits on $\sigma\cdot {\cal A}$ as a function of the
$q^*$ resonance mass after incorporation of systematic uncertainties.
Linear interpolations between test masses were used to determine where
the experimental bound intersected with a theoretical prediction to
yield a lower limit on allowed mass.  The corresponding observed 95\%
CL excited-quark mass exclusion region was found to be
$\MjjLowerExclusionTeV < m_{q^*} < \MjjLowerLimitMRSTTeV$~\TeVcc\
using MRST2007 PDFs in the ATLAS default MC09 tune.
Table~\ref{tab:PDFs} shows the results obtained using
CTEQ6L1~\cite{Pumplin:2002vw} and CTEQ5L~\cite{Lai:1999wy} PDF sets.
The variations in the observed limit associated with the error
eigenvectors of a CTEQ PDF set were found to be smaller than the
spread displayed in Table~\ref{tab:PDFs}.  The excluded regions were
$\sim\!30$~GeV greater when only statistical uncertainties were taken
into account.  The expected limits corresponding to the data sample
were computed using an analogous approach, but replacing the actual
data with pseudo-data generated by random fluctuations around the
smooth function described by fitting the data with Eqn.~\ref{eq:f};
these are shown in Fig.~\ref{fig:ExclusionWithStatSystematics}, with a
resulting expected $q^*$ mass exclusion region of
$\MjjLowerExclusionTeV < m_{q^*} <
\MjjLowerLimitMRSTEXPECTEDTeV$~\TeVcc\ using MRST2007 PDFs.  As
indicated in Table~\ref{tab:PDFs}, the two other PDF sets yielded
similar results, with expected exclusion regions extending to near
1~TeV.  An indication of the dependence of the $m_{q^*}$ limits on the
theoretical prediction for the $q^*$ signal was obtained by
simultaneously varying both the renormalization and factorization
scales by factors of 0.5 and 2, which was tantamount to modifying the
predicted cross section by approximately $\pm 20\%$; this changed the
observed MRST2007 limit of \MjjLowerLimitMRSTTeV~TeV to
\MjjLowerLimitMRSThalfmuRmuFTeV~TeV and
\MjjLowerLimitMRSTtwicemuRmuFTeV~TeV, respectively.

\begin{table*}
\caption[Mjj exclusion limits]{The 95\% CL lower limits on the allowed $q^*$ mass obtained
using different PDF sets.\label{tab:PDFs}\\}
\begin{ruledtabular}
\begin{tabular}{llccc}
~ & ~ & \multicolumn{2}{c}{Observed Mass Limit [TeV]} & Expected Mass Limit [TeV] \\
MC Tune & PDF Set  & Stat.~$\oplus$~Syst. & Stat.~only  & Stat.~$\oplus$~Syst. \\ \hline
MC09~\cite{ATLAS_MC09} & MRST2007~\cite{Sherstnev:2007nd} &
  \MjjLowerLimitMRSTTeV & \MjjLowerLimitStatOnlyMRSTTeV & \MjjLowerLimitMRSTEXPECTEDTeV \\
MC09$^\prime$~\footnote{The MC09$^\prime$ tune is identical to MC09 
      except for the {\sc pythia}~\cite{pythia} parameter PARP(82)$=2.1$ and use of the
      CTEQ6L1 PDF set.} &
  CTEQ6L1~\cite{Pumplin:2002vw} &
  \MjjLowerLimitCTEQsixLoneTeV & \MjjLowerLimitStatOnlyCTEQsixLoneTeV &
  \MjjLowerLimitCTEQsixLoneEXPECTEDTeV \\
Perugia0~\cite{Skands:2009zm} & CTEQ5L~\cite{Lai:1999wy} &
  \MjjLowerLimitCTEQfiveLTeV & \MjjLowerLimitStatOnlyCTEQfiveLTeV &
  \MjjLowerLimitCTEQfiveLEXPECTEDTeV \\
\end{tabular}
\end{ruledtabular}
\end{table*}

In conclusion, a model-independent search for new heavy particles
manifested as mass resonances in dijet final states was
conducted using a \integLumi~sample of 7~TeV proton-proton collisions
produced by the LHC and recorded by the ATLAS detector.  No evidence
of a resonance structure was found and upper limits at the 95\% CL
were set on the products of cross section and signal acceptance for
hypothetical new $q^*$ particles decaying to dijets.  These data
exclude at the 95\% CL excited-quark masses from the lower edge of the
search region, \MjjLowerExclusionTeV~\TeVcc, to
\MjjLowerLimitMRSTTeV~\TeVcc\ for a standard set of model parameters
and using the ATLAS default MC09 tune~\cite{ATLAS_MC09}.  This result
extends the reach of previous experiments and constitutes the first
exclusion of physics beyond the Standard Model by the ATLAS
experiment.  In the future, such searches will be extended to exclude
or discover additional hypothetical particles over greater mass
ranges.

\input{Acknowledgement-20apr10-collisions-modified.tex}
\bibliography{resonancePRL}



\begin{widetext}
\input{atlas_authlist_final_nonrevtex_20100902.tex}
\end{widetext}

\end{document}

%% file: Acknowledgement-20apr10-collisions-modified.tex


\section{Acknowledgements}



We deeply thank everybody at CERN involved in operating the LHC
in such a superb way during this initial high-energy data-taking
period. We acknowledge equally warmly all the technical and
administrative staff in the collaborating institutions without
whom ATLAS could not be operated so efficiently.

We acknowledge the support of ANPCyT, Argentina; Yerevan Physics
Institute, Armenia; ARC and DEST, Australia; Bundesministerium f\"ur
Wissenschaft und Forschung, Austria; National Academy of Sciences of
Azerbaijan; State Committee on Science \& Technologies of the Republic
of Belarus; CNPq and FINEP, Brazil; NSERC, NRC, and CFI, Canada; CERN;
CONICYT, Chile; NSFC, China; COLCIENCIAS, Colombia; Ministry of
Education, Youth and Sports of the Czech Republic, Ministry of
Industry and Trade of the Czech Republic, and Committee for
Collaboration of the Czech Republic with CERN; Danish Natural Science
Research Council and the Lundbeck Foundation; European Commission,
through the ARTEMIS Research Training Network; IN2P3-CNRS and
CEA-DSM/IRFU, France; Georgian Academy of Sciences; BMBF, DFG, HGF and
MPG, Germany; Ministry of Education and Religion, through the EPEAEK
program PYTHAGORAS II and GSRT, Greece; ISF, MINERVA, GIF, DIP, and
Benoziyo Center, Israel; INFN, Italy; MEXT, Japan; CNRST, Morocco; FOM
and NWO, Netherlands; The Research Council of Norway; Ministry of
Science and Higher Education, Poland; GRICES and FCT, Portugal;
Ministry of Education and Research, Romania; Ministry of Education and
Science of the Russian Federation and State Atomic Energy Corporation
ROSATOM; JINR; Ministry of Science, Serbia; Department of
International Science and Technology Cooperation, Ministry of
Education of the Slovak Republic; Slovenian Research Agency, Ministry
of Higher Education, Science and Technology, Slovenia; Ministerio de
Educaci\'{o}n y Ciencia, Spain; The Swedish Research Council, The Knut
and Alice Wallenberg Foundation, Sweden; State Secretariat for
Education and Science, Swiss National Science Foundation, and Cantons
of Bern and Geneva, Switzerland; National Science Council, Taiwan;
TAEK, Turkey; The Science and Technology Facilities Council and The
Leverhulme Trust, United Kingdom; DOE and NSF, United States of
America.

%% file: atlas_authlist_final_nonrevtex_20100902.tex
\begin{flushleft}
{\Large The ATLAS Collaboration}

\bigskip

G.~Aad$^{\rm 48}$,
B.~Abbott$^{\rm 111}$,
J.~Abdallah$^{\rm 11}$,
A.A.~Abdelalim$^{\rm 49}$,
A.~Abdesselam$^{\rm 118}$,
O.~Abdinov$^{\rm 10}$,
B.~Abi$^{\rm 112}$,
M.~Abolins$^{\rm 88}$,
H.~Abramowicz$^{\rm 153}$,
H.~Abreu$^{\rm 115}$
% ...
\bigskip

% ...
$^{10}$ Institute of Physics, Azerbaijan Academy of Sciences, H. Javid Avenue 33, AZ 143 Baku, Azerbaijan\\
$^{11}$ Institut de F\'isica d'Altes Energies, IFAE, Edifici Cn, Universitat Aut\`onoma  de Barcelona,  ES - 08193 Bellaterra (Barcelona), Spain\\
% ...
$^{48}$ Albert-Ludwigs-Universit\"{a}t, Fakult\"{a}t f\"{u}r Mathematik und Physik, Hermann-Herder Str. 3, D - 79104 Freiburg i.Br., Germany\\
$^{49}$ Universit\'e de Gen\`eve, Section de Physique, 24 rue Ernest Ansermet, CH - 1211 Geneve 4, Switzerland\\
% ...
$^{88}$ Michigan State University, Department of Physics and Astronomy, High Energy Physics Group, East Lansing, MI 48824-2320, United States of America\\
% ...
$^{111}$ University of Oklahoma, Homer L. Dodge Department of Physics and Astronomy, 440 West Brooks, Room 100, Norman, OK 73019-0225, United States of America\\
$^{112}$ Oklahoma State University, Department of Physics, 145 Physical Sciences Building, Stillwater, OK 74078-3072, United States of America\\
% ...
$^{115}$ LAL, Univ. Paris-Sud, IN2P3/CNRS, Orsay, France\\
% ...
$^{118}$ Oxford University, Department of Physics, Denys Wilkinson Building, Keble Road, Oxford OX1 3RH, United Kingdom\\
% ...
$^{153}$ Tel Aviv University, Raymond and Beverly Sackler School of Physics and Astronomy, Ramat Aviv, IL - Tel Aviv 69978, Israel\\
% ...
\end{flushleft}
